\documentclass[prl,aps,twocolumn,superscriptaddress]{revtex4}
\usepackage{graphicx}
\usepackage{amssymb, amsmath, color, rotating,bm, mathtools}
\usepackage{color}
\usepackage{float}

\newcommand{\kk}{\mathbf{k}}
\newcommand{\rrho}{\boldsymbol{\rho}}
\usepackage{braket}

\begin{document}

\title{Anomalous supersolidity in a weakly interacting dipolar Bose mixture on a square lattice}

 \author{Ryan M. Wilson}
 \affiliation{Department of Physics, The United States Naval Academy, Annapolis, MD 21402, USA}
 \author{Wilbur E. Shirley}
 \affiliation{Condensed Matter Theory Center and Joint Quantum Institute, Department of Physics, University of Maryland, College Park, Maryland 20742-4111 USA}
 \affiliation{California Institute of Technology, Pasadena, CA 91125  USA}
 \email{rwilson@usna.edu}
 \author{Stefan S. Natu}
 \affiliation{Condensed Matter Theory Center and Joint Quantum Institute, Department of Physics, University of Maryland, College Park, Maryland 20742-4111 USA}

\begin{abstract}
We calculate the mean-field phase diagram of a zero-temperature, binary Bose mixture on a square optical lattice, where one species possesses a non-negligible dipole moment.  Remarkably, this system exhibits supersolidity for anomalously weak dipolar interaction strengths, which are readily accessible with current experimental capabilities. The supersolid phases are robust, in that they occupy large regions in the parameter space.  Further,  we identify a  first-order quantum phase transition between supersolid and superfluid phases.   Our results demonstrate the rich features of the dipolar Bose mixture, and suggest that this system is well-suited for exploring supersolidity in the experimental setting.
\end{abstract}
\maketitle

\emph{Introduction}-- The physics of emergent, competing orders is central to the rich phenomenology of many condensed matter systems, such as  high-$T_\mathrm{c}$ superconductors and frustrated magnets.  Recently, exciting developments in the cooling and trapping of magnetic atoms~\cite{Griesmaier05,Lahaye07,Lahaye09,Lu11,Lu12,Aikawa12,Bismut12,Baranov12} and diatomic molecules~\cite{Ni08,Deiglmayr08,Aikawa09,Carr09,Takekoshi14} offer promise that the physics of competing orders will be accessible in exceptionally clean, controllable forms of \emph{synthetic} quantum matter.  One striking example is the predicted supersolid phase of strongly dipolar bosons loaded in an optical lattice, where the system simultaneously exhibits crystalline order and superfluidity.  
Indeed, checkerboard and stripe supersolids are predicted to emerge in dipolar lattice systems, in addition to a variety of structured insulating phases~\cite{Goral02,Kovrizhin05,Scarola05,Yi07,Menotti07,Danshita09,Burnell09,Cinto10,Trefzger10,Pollet10,Iskin11,Trefzger11,Fellows11,Ohgoe12,Yamamoto12,Lu15}.   
The study of supersolidity predates experiments with ultracold atoms~\cite{Chester70}, and was first proposed as a potential manifestation of solidity in superfluid $^4$He~\cite{Chester70,Leggett70}.  Despite significant, long-standing interest in this phase and controversy over its existence~\cite{Kim04,Kim06,Todoshchenko07,Choi10,Kim12}, a supersolid ground state has yet to be observed in an experimental setting.

\begin{figure}[b!]
\includegraphics[width=.7\columnwidth]{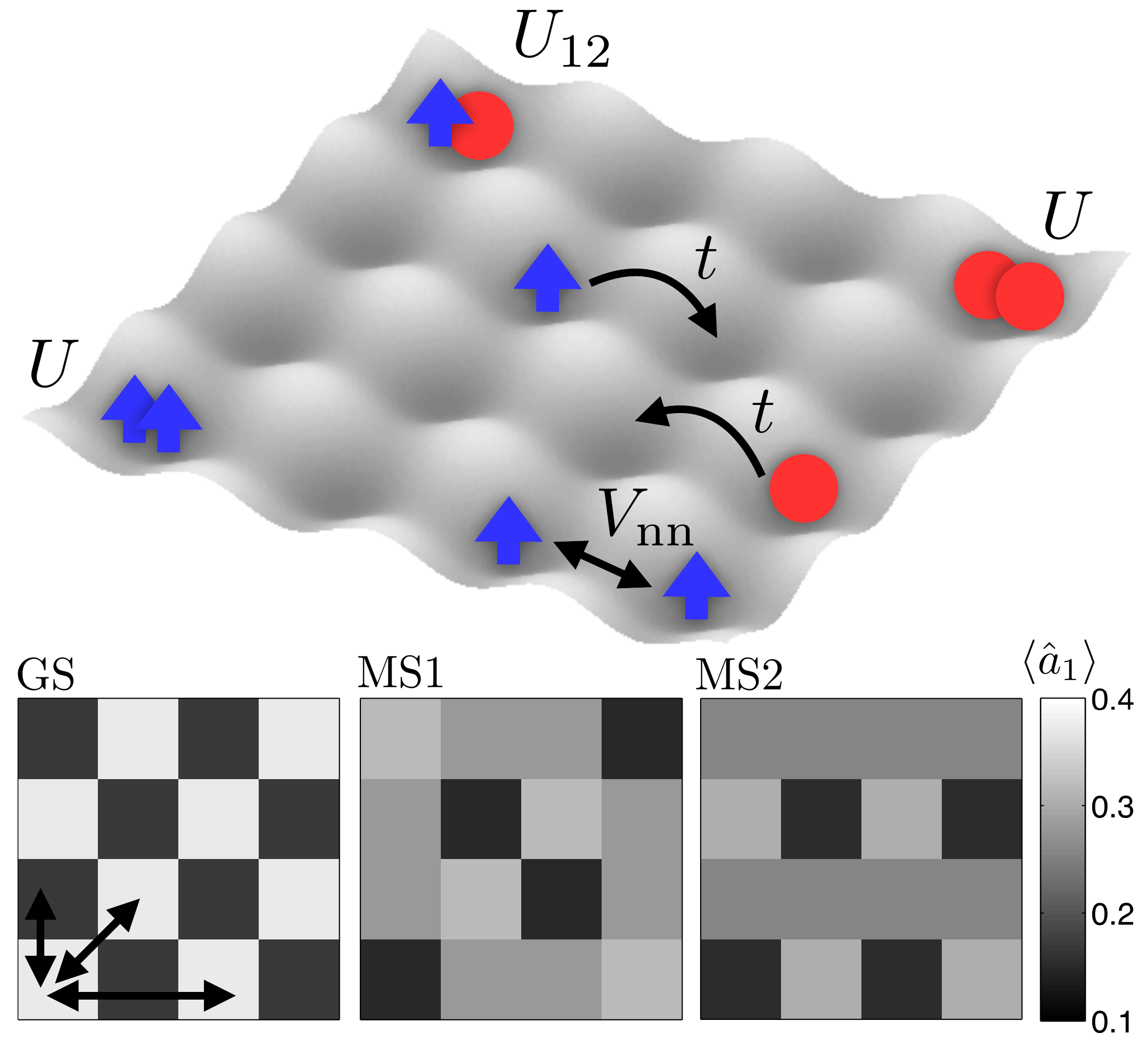}
\caption{\label{schematic} Schematic showing a mixture of equal-mass dipolar (blue arrows) and non-dipolar (red circles) bosons in a square optical lattice.  The local intraspecies interactions $U$ are assumed to be species independent, $U_{12}$ is the local interspecies interaction strength, and $t$ is the hopping rate.
Nearest neighbor (nn), next nearest neighbor, and next-next nearest neighbor dipolar interactions are shown schematically by the black arrows in the bottom panel.  This panel shows the spatial dependence of the superfluid order parameter $\langle \hat{a}_1 \rangle$ for the dipolar species in the supersolid regime (see text).  The ground state (GS) is a checkerboard supersolid.  Two nearly degenerate, metastable excited states are depicted as MS1 and MS2. } 
\end{figure}

 In the context of ultracold atoms in optical lattices, it is often the case that the long-range dipolar interactions, which are responsible for discrete translational symmetry breaking and  the formation of crystalline order~\cite{Otterlo94,Otterlo95,Batrouni95,Baumann10}, are typically very weak, being easily overwhelmed by atomic motion (hopping), repulsive local interactions, and finite temperature, which favor spatially uniform phases.   Other proposals suggest that stronger effective dipolar interactions can be achieved by using large densities, though local interactions can easily destroy supersolid order in this semiclassical regime~\cite{Buhler11}.   Here, we show that the challenge posed by dipolar interactions that are weak compared to the other energy scales in the system  is readily overcome by working with a \textit{binary} mixture of bosons, where a non-dipolar species is cospatial and interacting with the dipolar system~\cite{Wilson12,Shirley14}.   
 
 We calculate the zero-temperature ($T=0$) phase diagram for this system in the experimentally relevant square lattice geometry, using a site-decoupled Gutzwiller mean-field method~\cite{Rokhsar91}.  
 A key parameter in our theory is the local interspecies interaction strength, which encourages translational symmetry breaking, and thus  the formation of a supersolid.  Our results demonstrate that the supersolid phase occupies an anomalously large region in the phase diagram compared to scalar dipolar Bose gases, and persists even for very weak dipolar interaction strengths. 
  Thus, we propose that the binary dipolar Bose mixture is a novel system for exploring the interplay between superfluidity and crystalline order in the cold atoms context, and is a promising candidate for the experimental realization of supersolidity.

\textit{Dipolar mixture}-- 
The tight-binding Hamiltonian for an equal-mass mixture of dipolar ($\sigma=1$) and non-dipolar ($\sigma=2$) bosonic atoms reads (see Fig.~\ref{schematic}) 
\begin{eqnarray}\label{ham}
\hat{H} &=& -t\sum_{\langle ij\rangle\sigma}(\hat{a}^\dag_{\sigma i}\hat{a}_{\sigma j}  + \text{h.c.}) - \sum_{i} \mu_\sigma \hat{n}_{\sigma i} \nonumber \\ &+& 
\sum_{i , \sigma \sigma^\prime} \frac{U_{\sigma \sigma^\prime}}{2} \hat{n}_{\sigma i} \left( \hat{n}_{\sigma^\prime i}  - \delta_{\sigma \sigma^\prime}  \right) +\sum_{i<j}V_{ij}\hat{n}_{1 i}\hat{n}_{1 j}. 
\end{eqnarray}
Here, $\hat{a}_{i\sigma}$ ($\hat{a}^\dagger_{i \sigma}$) annihilates (creates) a boson of species $\sigma$ at site $i$, $\hat{n}_{\sigma i} = \hat{a}_{\sigma i}^\dagger \hat{a}_{\sigma i}$ is the number operator for species $\sigma$ on site $i$, $\mu_{\sigma}$ is the global chemical potential for species $\sigma$, and $U_{12}=U_{21}$ is the local interspecies interaction strength.  For simplicity, we take the local intraspecies interaction strengths to be equal, setting $U_{11} = U_{22} \equiv U$.  

We consider dipoles aligned perpendicular to the lattice plane,  yielding purely repulsive, isotropic dipolar interaction couplings  $V_{ij} = d^{2}/|\rrho_{i}-\rrho_{j}|^{3}$, where $d$ denotes the dipole moment, and $\rrho_{j}$ are the positions of the dipolar atoms on the square 2D lattice with spacing $a$; we rescale all lengths by $a$.  We denote the nearest neighbor dipolar interaction couplings as $V_\mathrm{nn}$. Although the dipolar interactions are long range, they are significantly weaker than typical local interaction strengths for the magnetic atoms Cr, Dy, and Er \cite{Lahaye09}.  For example, the nearest-neighbor dipolar coupling for $^{168}$Er ($d_\mathrm{Er} = 7 \, \mu_\mathrm{B}$ where $\mu_\mathrm{B}$ is the Bohr magneton) in a square lattice with spacing $a = 266 \, \mathrm{nm}$ is $V_\mathrm{nn} = d_\mathrm{Er}^2 / a^3 \simeq h \times 34 \, \mathrm{Hz}$~\cite{BaierarXiv15}, whereas $U$ is typically many kHz.  $V_\mathrm{nn}$ should be about twice as large for Dy atoms under equivalent conditions.  Thus, we can reasonably expect $0.01 \lesssim V_\mathrm{nn} / U \lesssim 0.1$ for atomic systems, with longer range interactions being further suppressed by the $1/\rho^3$ scaling.  A key distinction of our work is that we identify supersolid phases in this range of weak dipolar interaction strengths for a dipolar Bose mixture.  
In contrast, dipolar interaction strengths of $V_\mathrm{nn} / U \gtrsim 0.5$ are required to produce supersolidity in a single species dipolar system~\cite{Danshita09,Iskin11}.
Much stronger dipolar interactions can be achieved with diatomic, heteronuclear molecules, which are being actively pursued experimentally~\cite{Carr09}.  However, it is difficult to achieve large phase space densities with diatomic molecules, and collisional losses impose considerable constraints on their utility~\cite{Ospelkaus10,Ni10,Mayle13}.  

We employ a Gutzwiller mean-field theory to obtain the $T=0$  phase diagram of Eq.~(\ref{ham}), and introduce a spatially varying superfluid order parameter $\langle \hat{a}_{\sigma i} \rangle$ for each species $\sigma$~\cite{Rokhsar91}. 
Throughout, we find ground states with either uniform or checkerboard spatial order.  When dipolar interactions beyond nearest-neighbor are considered, our method unveils a manifold of nearly-degenerate, metastable excited states with supersolid ordering at multiple wave vectors~\cite{Menotti07}.  Examples are depicted schematically in Fig.~\ref{schematic} for a $4 \times 4$ unit cell with period boundary conditions (MS1 \& MS2), and parameters $t/U = 0.03$, $\mu_1/ U = \mu_2 / U = 2.5$, $U_{12} / U = 0.9$, and $V_\mathrm{nn}/U = 0.1$.  The metastable states are gapped from the ground state by an energy proportional to the dipolar interaction strength.  Because uniform and checkerboard orders possess an $AB$ sublattice symmetry, we consider only the nearest-neighbor part of the dipolar interactions and specialize to a $2 \times 2$ unit cell with periodic boundary conditions, and focus on the ground state phases only.   Additionally, we fix the chemical potentials to be equal, $\mu_{1} = \mu_{2} \equiv \mu$.  We vary $\mu$ as a free parameter in the theory, which controls the total atom number $N = \sum_{\sigma i} n_{\sigma i}$, where $n_{\sigma i} = \langle \hat{n}_{\sigma i} \rangle$.  Because the dipolar interactions break the interspecies symmetry of the system, this choice produces a number imbalance that scales with $V_\mathrm{nn}$.  The imbalance remains relatively small, however, for the weak dipolar interactions we consider here.  We focus on the regime $0 < U_{12} < U$, which discourages spatial demixing of the species. 

\begin{figure}[t!]
\includegraphics[width=.8\columnwidth]{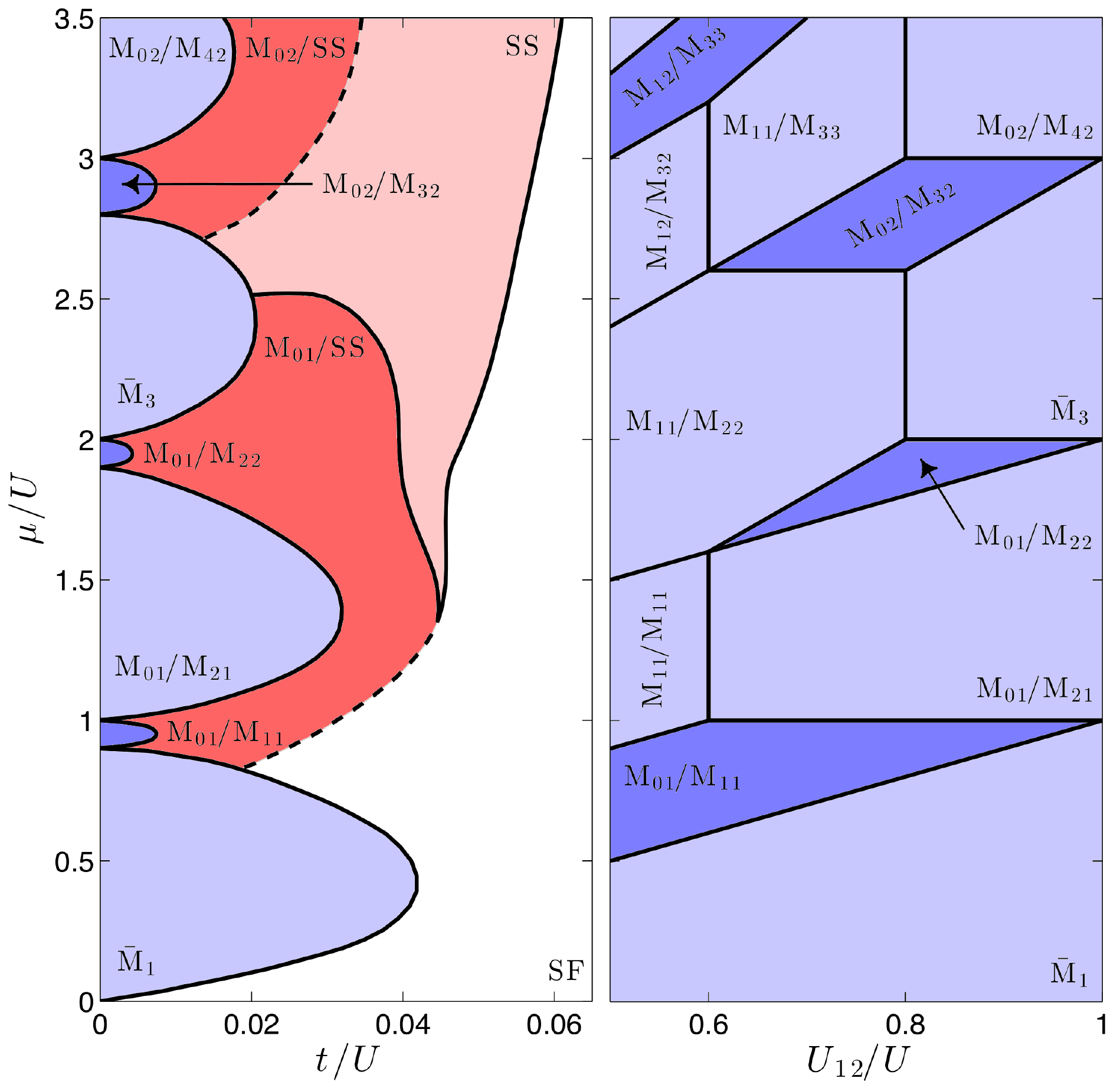} \hfill
\caption{\label{pd} (Left) Ground state phase diagram for $U_{12}/U = 0.9$ and $V_\mathrm{nn}/U = 0.1$, exhibiting a large supersolid region.  The light blue lobes correspond to Mott insulators with spatially uniform total density, and checkerboard order in the individual species.  The dark blue lobes correspond to Mott insulators with checkerboard structure in the total density.  The red regions correspond to $M_{0 n_{1B}}/$SS phases,  the pink regions correspond to SS phases, and the white region corresponds to a spatially uniform superfluid (SF).
The solid (dashed) black lines show second-order (first-order) phase transitions. (Right) Phase diagram at $t=0$, varying $U_{12}$.  All phases are Mott insulators, with coloring equivalent to that in the left panel.   (See text for details.) }
\end{figure}


\textit{Results}-- In the left panel of Fig.~\ref{pd}, we present the phase diagram obtained for $U_{12}/U = 0.9$ and $V_\mathrm{nn}/U = 0.1$ as a function of $t$ and $\mu$. For larger $t$, corresponding to shallower lattice depths, the system is a spatially uniform superfluid (SF), characterized by non-zero values of the $k$-space superfluid order parameters $ \tilde{\alpha}_\sigma (\kk) = \sum_{i}e^{i  \kk \cdot \rrho_i} \langle \hat{a}_{i \sigma} \rangle$ at $\kk = (k_x , k_y) =  (0, 0)$.  As $t$ is decreased, the superfluid order parameter(s) acquire weight at $\kk = (\pi,\pi)$, signifying the transition to a checkerboard supersolid phase.  Supersolidity can manifest in two distinct ways in this bosonic mixture: both species can exhibit supersolid order (the SS phase), or the dipolar species can transition directly from a SF to a checkerboard Mott insulator while the non-dipolar species remains superfluid.  In the latter case, the dipolar species forms an effective checkerboard potential for the non-dipolar species due to their mutually repulsive interactions, which results in superfluidity with density-wave order,  or supersolidity, for the non-dipolar species.  We denote the Mott insulator phases by $\mathrm{M}_{n_A n_B}$, where $n_{A(B)}$ are the integer occupations of the $A(B)$ sublattice sites. 
This phase diagram possesses a tri-critical point 
between the $\bar{\mathrm{M}}_3$, SS, and M$_{01}/$SS phases, though we do not study this point in detail here.

The SF-SS transition occurs at larger densities, for $\mu/U \gtrsim 1.5$, and is second-order, indicated by the solid black line in Fig.~\ref{pd}.  In contrast, the transition to the M$_{01}$/SS phase, where the dipolar species is in the M$_{01}$ phase and the non-dipolar species is SS, occurs at smaller densities and is strongly first-order, indicated by the dashed black line in this figure.  We note that first-order transitions between purely insulating and SF phases were predicted in previous theoretical studies of non-dipolar Bose mixtures~\cite{Kuklov04,Isacsson05,Yamamoto13}, and a first-order SF-SS transition was predicted for hard-core dipolar bosons on a triangular lattice~\cite{Yamamoto12}.  The presence of a first-order superfluid-supersolid phase transition for \emph{weak} dipolar interactions is a new feature of the system we consider here.

\begin{figure}[t!]
\includegraphics[width=.7\columnwidth]{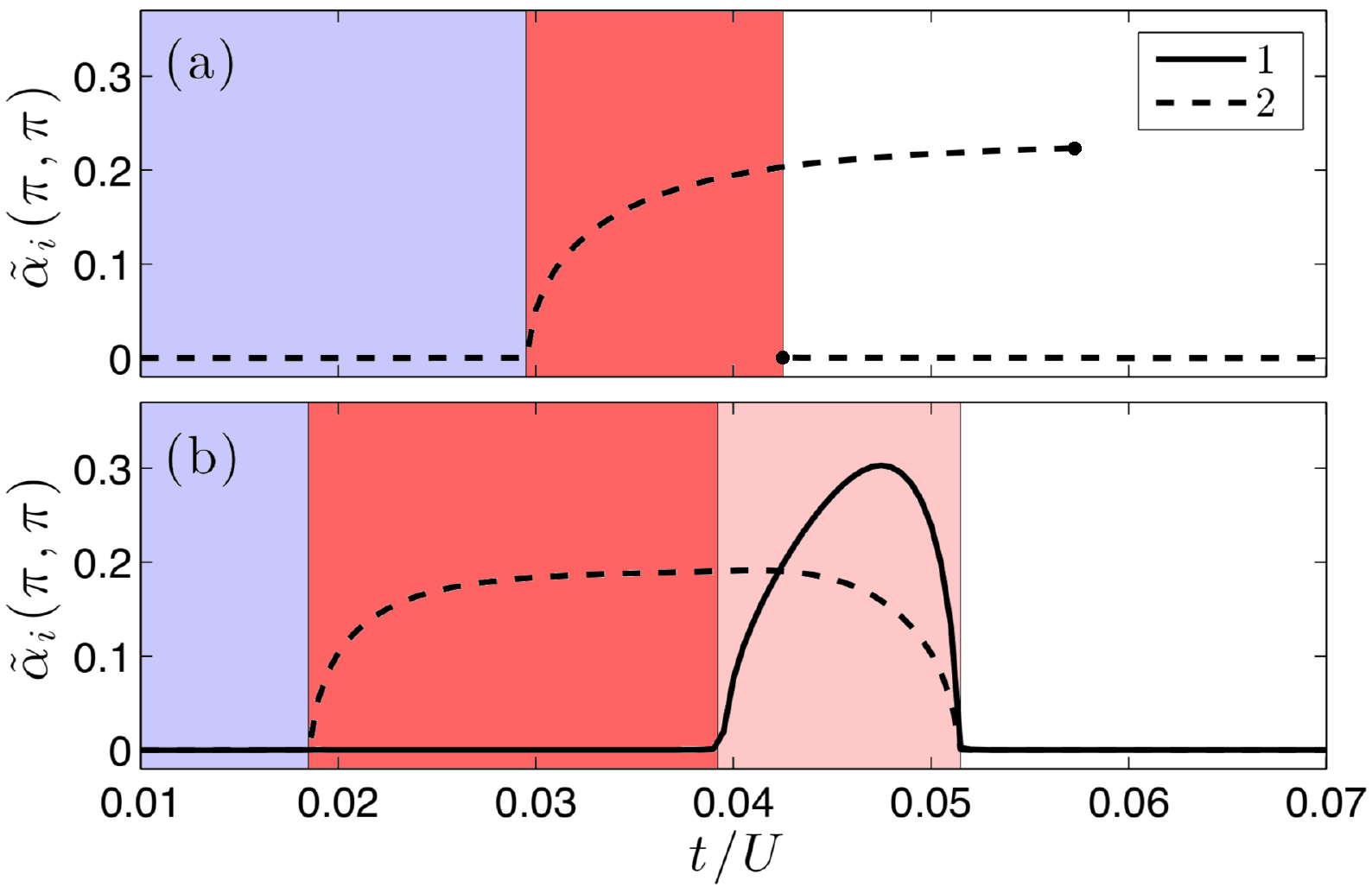} \hfill
\caption{\label{ssorder}  (a) Supersolid order parameter as a function of $t/U$ for $V_\mathrm{nn} = 0.1U$, $U_{12} = 0.9U$, and $\mu = 1.25U$.  (b) Same as top, except for $\mu = 2.25U$.  The blue regions correspond to Mott insulators, the red regions correspond to M$_{0 n_{1B}}/$SS phases, the pink corresponds to a SS, and the white regions correspond to uniform superfluids (SF).  All transitions are second-order except the transition to a SF in (a), which is strongly first-order.  The double-valued order parameter is characteristic of hysteresis at a first-order transition.  }
\end{figure}

We demonstrate the first-order nature of the M$_{01} / $SS to SF phase transition in Fig.~\ref{ssorder}, where the $k$-space superfluid order parameters at $\kk = (\pi, \pi)$ are shown for $\mu / U = 1.25$ in panel (a) and  $\mu / U = 2.25$ in panel (b), corresponding to horizontal cuts across the left panel of Fig.~\ref{pd} (with $U_{12} / U = 0.9$ and $V_\mathrm{nn} / U= 0.1$).
In panel (b), the order parameters change continuously across the transition from a M$_{03}/$M$_{30}$ insulator, through the M$_{01}/$SS and SS phases, to a SF phase.  For $\mu / U = 1.25$, $\tilde{\alpha}_1(\pi,\pi)=0$ for all values of $t/U$, so only $\tilde{\alpha}_2(\pi,\pi)$ is shown in panel (a).  Here, the transition from a M$_{01}$/M$_{21}$ insulator to a M$_{01}/$SS supersolid is second-order, while the transition to a SF is clearly discontinuous, and first-order. By smoothly following our ground state solutions from either side of the transition region, we find that $\tilde{\alpha}_2(\pi,\pi) $ is multivalued for $0.04 \lesssim t/U \lesssim 0.06$; this is indicative of hysteresis, which is a feature of first-order phase transitions. 
Notice that a first-order transition also exists between the M$_{02}$/SS and SS phases, shown in Fig.~\ref{pd}, though the hysteresis area of this transition is notably smaller.  

For sufficiently small $t$, corresponding to deeper lattices, both species enter Mott insulating phases, indicated by the blue lobes in the left panel of Fig.~\ref{pd}.  The larger, light blue lobes correspond to insulating phases with a spatially uniform density of $n$ atoms per site. 
 When $n$ is even, the individual species form checkerboard Mott insulators, where $n_{1A}+n_{2A} = n_{1B}+n_{2B} = n$.  When $n$ is odd, degeneracies exist between insulating phases with different combinations of $n_{\sigma A}$ and $n_{\sigma B}$.   For example, the first Mott lobe in Fig.~\ref{pd} corresponds to $n=1$, and has a degeneracy between the M$_{00}$/M$_{11}$ and M$_{01}$/M$_{10}$ phases.  The third Mott lobe corresponds to $n=3$, and has a degeneracy between the M$_{01}$/M$_{32}$ and M$_{02}$/M$_{31}$ phases.   These phases are labeled $\bar{\mathrm{M}}_1$ and $\bar{\mathrm{M}}_3$, respectively, in Fig.~\ref{pd}.  We note that this degeneracy is a consequence of our choice $\mu_1 = \mu_2$, and is broken if we instead enforce equal total atom number, $N_1 = N_2$.   Interestingly, the Mott lobes with uniform $n$ are separated by smaller lobes, wherein the dipolar species forms a checkerboard insulator and the non-dipolar species forms a uniform insulator, resulting in a Mott insulator phase with checkerboard ordering in the \emph{total} density.

To explore this further, we calculate the $t=0$ phase diagram as a function of $\mu$ and $U_{12}$ for $V_\mathrm{nn} / U = 0.1$, shown in the right panel of Fig.~\ref{pd}.  Unlike the single species dipolar system, the Mott physics of the dipolar mixture is quite rich, and exhibits an interesting array of insulating phases.  In particular, the diagram shows that Mott insulating phases with checkerboard order in the total density (dark blue regions) are sizable for smaller $U_{12}$, and shrink linearly as $U_{12} \rightarrow U$.   For $t>0$, these lobes melt into M$_{0 n_{1B}}$/SS supersolid phases.  At exactly $U_{12} = U$, these phases vanish, and the system only supports insulting phases with uniform total density.  

\begin{figure}[t!]
\includegraphics[width=.7\columnwidth]{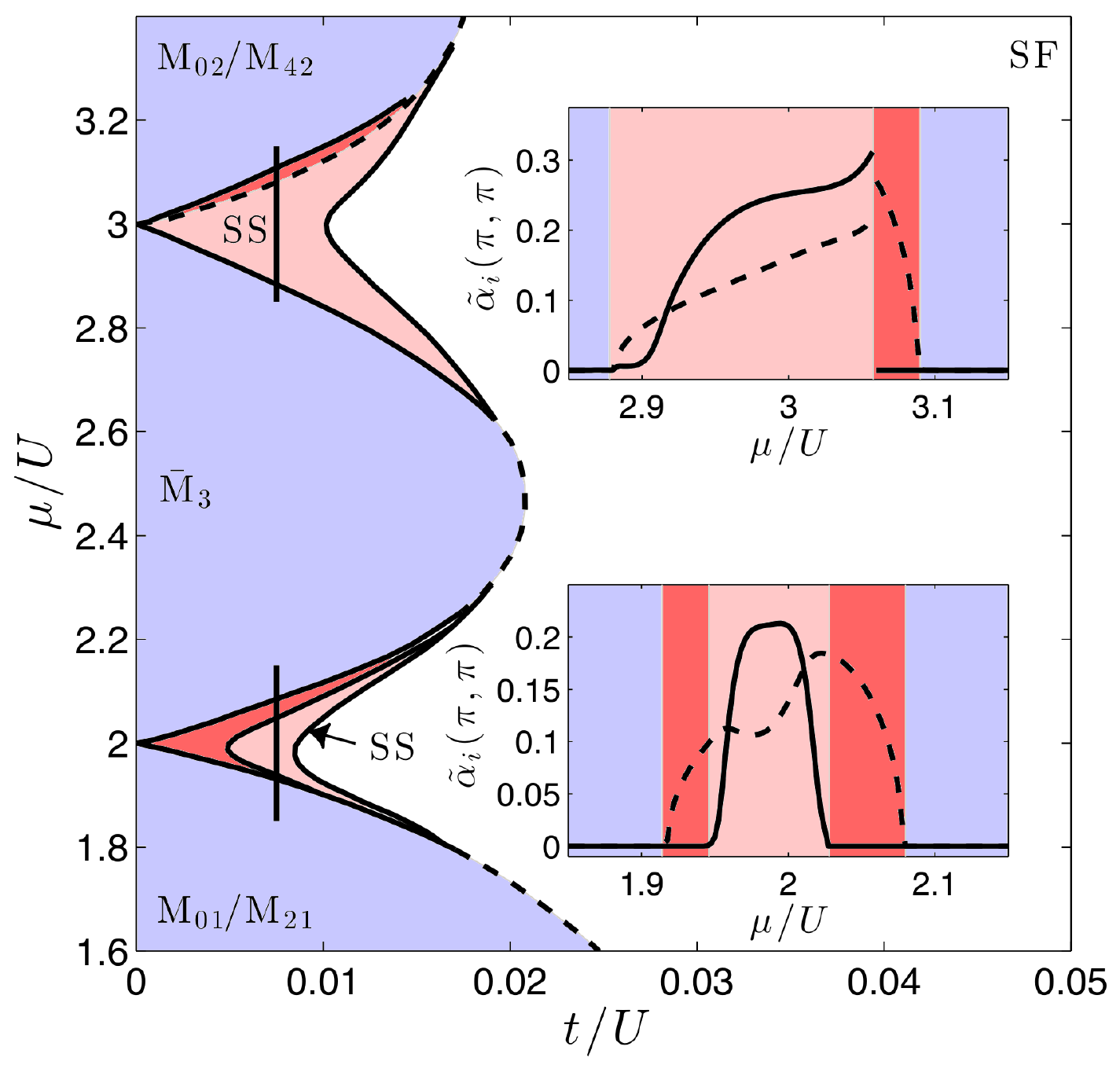}
\caption{\label{smallV}  Ground state phase diagram for $U_{12}/U = 0.99$ and $V_\mathrm{nn}/U = 0.01$, exhibiting supersolidity for very weak dipolar interactions.  The light blue lobes correspond to Mott insulators with spatially uniform total density, the red regions correspond to M$_{0 n_{1B}}/$SS phases, the pink regions correspond to SS phases, and the white region corresponds to a spatially uniform superfluid.  The solid (dashed) black lines in the main panel show second-order (first-order) phase transitions.  The lower (upper) inset shows the supersolid order parameters for $t/U = 0.8$ at a cut near $\mu/U = 3$ ($\mu / U = 2$).  In the insets, the solid line corresponds to the dipolar species ($\sigma = 1$) and the dashed line corresponds to the non-dipolar species ($\sigma = 2$).  The first-order phase transition is apparent as a discontinuity in the order parameter near $\mu/U \simeq 3.06$ in the upper inset.  }
\end{figure}

The tendency for ground states to acquire checkerboard density-wave order can be understood intuitively for this system, as this  minimizes the nearest neighbor contributions to the dipolar interaction energy in a square lattice geometry.
This ordering is preferred by the interspecies interactions ($U_{12} >0$), as well; the right panel of Fig.~\ref{pd} shows that all insulating phases at $U_{12}/U > 0.6$ have checkerboard ordering in the individual species.  The kinetic energy and local intraspecies interactions, however, prefer spatially uniform phases.  We thus expect supersolidity to vanish as the dipolar interactions weaken, unless $t$ is sufficiently small and $U_{12}$ is comparable to $U$.   This suggests that supersolidity may persist for very small dipolar interaction strengths, provided $t$ and  $1-U_{12}/U$ are sufficiently small.

In Fig.~\ref{smallV}, we plot the phase diagram for $U_{12}/U = 0.99$ and $V_\mathrm{nn}/U = 0.01$, corresponding to very weak dipolar interactions, as a function of $t$ and $\mu$. We note that very small M$_{01}/$M$_{22}$ and M$_{02}/$M$_{32}$ Mott lobes exist near $\mu/U = 2$ and $\mu/U = 3$ and $t\sim 0$, respectively, but are omitted from this diagram due to their vanishingly small size.  Strikingly, supersolid regions exist at small $t$, between adjacent checkboard Mott lobes, and  still occupy a significant region of the phase diagram.  The insets in Fig.~\ref{smallV} show the checkerboard supersolid order parameter $\tilde{\alpha}_i(\pi,\pi)$ for the dipolar species (solid black lines) and the non-dipolar species (dashed black lines) for $t/U = 0.8$. The lower inset corresponds to a cut near $\mu/U=2$, shown by the vertical black line to the left of the inset.  Here, the phase transitions from Mott insulator to supersolid are all continuous, and second-order.  The upper inset corresponds to a cut near $\mu/U=3$.  Here, the $\bar{M}_3$-SS transition is second-order, while the transition from M$_{02}$/SS to SS is discontinuous, and first-order; this is consistent with the phase diagram in Fig.~\ref{pd} for $V_\mathrm{nn} / U =0.1$.  Additionally, the  Mott insulator to SF transitions are first-order for larger $t$, as indicated by the dashed black lines near the tips of the Mott lobes.  This is consistent with the findings of Refs.~\cite{Kuklov04,Isacsson05,Yamamoto13}, where a first-order superfluid-insulator transition is predicted for a non-dipolar Bose mixture.

 Though our discussion has focused on interspecies interactions $0<U_{12}<U$, we note that supersolidity persists for $U_{12}>U$ as well.  We have performed analogous calculations to those described above, but with $\mu_1 > \mu_2$ chosen to balance the total particle number.  The Mott insulator states in this case are of the checkerboard form M$_{0n}/$M$_{n0}$ for all non-vanishing (nearest-neighbor) dipolar interaction strengths.  For $V_\mathrm{nn}/U = 0.01$ and $U_{12} / U = 1.01$, we find SS regions between the Mott lobes at small $t$, similar to those shown in Fig.~\ref{smallV}.  For $V_\mathrm{nn}/U = 0.1$ and $U_{12}  / U = 1.1$, we find a large SS region that extends to larger $t$, well beyond the Mott lobes.  While increasing $U_{12}$ will eventually lead to phase separation at finite $t$, the Mott lobes will possess checkerboard order for any $U_{12}>0$, and we thus expect SS regions to exist between these Mott lobes for sufficiently small $t$.


\textit{Discussion}--  In an experiment, the dipolar Bose mixture will inevitably have unequal masses, and thus species-dependent hopping.   We note that the supersolid regions span a large range of $t$ values, so supersolidity should persist for moderate differences in the species-dependent hopping rates.  Additionally, the presence of low-lying metastable states at energies $\sim V_\mathrm{nn}$ above the ground state suggests that very low temperatures will be necessary to realize pure checkerboard ground states.  For the Er example discussed above, with $V_\mathrm{nn} \simeq h \times 34 \,\mathrm{Hz}$, temperatures on the order of a few nK are sufficient to discourage population of these metastable states.  Still, we note that these excited states are supersolid in nature, and should permit superfluid transport and show signatures of crystalline order in Bragg spectroscopy~\cite{Stenger99,Hart15} at super-critical temperatures.  In previous theoretical studies of single species dipolar systems, beyond mean-field  effects were found to enhance the Mott lobes, and only slightly diminish the supersolid regions~\cite{Capogrosso08,Capogrosso10,Trefzger11}.  We therefore expect the supersolid phases we find here to be robust against quantum effects at intermediate densities, even for $V_\mathrm{nn}/U = 0.01$ (c.f.~Fig.~\ref{smallV}).

The subject of supersolidity in condensed matter systems has a rich history, and remains an active area of research~\cite{Kadau15arXiv}.  
Despite evidence of supersolid phases in non-equilibrium systems~\cite{Baumann10}, the observation of this phase as a ground state remains an open problem. Lattice analogues of supersolid phases in long-range interacting systems are a promising avenue to explore this physics, as the lattice naturally enhances correlations by suppressing the kinetic energy, while the long range interactions introduce a natural length scale for breaking discrete spatial symmetry.
Here, we have shown that a dipolar Bose mixture on a square lattice is a promising candidate for realizing supersolid ground states, even in the presence of anomalously weak dipolar interaction strengths.



\textit{Acknowledgements}-- We acknowledge B. M. Anderson for helpful conversations in the early stages of this work.   RW acknowledges partial support from the Office of Naval Research under Grant No.~N00014115WX01372, and from the National Science Foundation under Grant No.~PHY-1516421.  WS acknowledges support from a JQI-PFC Seed Grant. SN thanks the LPS-CMTC, LPS-MPO-CMTC, JQI-NSF-PFC, and ARO-MURI for support.


\begin{thebibliography}{56}
\expandafter\ifx\csname natexlab\endcsname\relax\def\natexlab#1{#1}\fi
\expandafter\ifx\csname bibnamefont\endcsname\relax
  \def\bibnamefont#1{#1}\fi
\expandafter\ifx\csname bibfnamefont\endcsname\relax
  \def\bibfnamefont#1{#1}\fi
\expandafter\ifx\csname citenamefont\endcsname\relax
  \def\citenamefont#1{#1}\fi
\expandafter\ifx\csname url\endcsname\relax
  \def\url#1{\texttt{#1}}\fi
\expandafter\ifx\csname urlprefix\endcsname\relax\def\urlprefix{URL }\fi
\providecommand{\bibinfo}[2]{#2}
\providecommand{\eprint}[2][]{\url{#2}}

\bibitem[{\citenamefont{Griesmaier et~al.}(2005)\citenamefont{Griesmaier,
  Werner, Hensler, Stuhler, and Pfau}}]{Griesmaier05}
\bibinfo{author}{\bibfnamefont{A.}~\bibnamefont{Griesmaier}},
  \bibinfo{author}{\bibfnamefont{J.}~\bibnamefont{Werner}},
  \bibinfo{author}{\bibfnamefont{S.}~\bibnamefont{Hensler}},
  \bibinfo{author}{\bibfnamefont{J.}~\bibnamefont{Stuhler}}, \bibnamefont{and}
  \bibinfo{author}{\bibfnamefont{T.}~\bibnamefont{Pfau}},
  \bibinfo{journal}{Phys. Rev. Lett.} \textbf{\bibinfo{volume}{94}},
  \bibinfo{pages}{160401} (\bibinfo{year}{2005}).

\bibitem[{\citenamefont{Lahaye et~al.}(2007)\citenamefont{Lahaye, Koch,
  Fr\"ohlich, Fattori, Metz, Griesmaier, Giovanazzi, and Pfau}}]{Lahaye07}
\bibinfo{author}{\bibfnamefont{T.}~\bibnamefont{Lahaye}},
  \bibinfo{author}{\bibfnamefont{T.}~\bibnamefont{Koch}},
  \bibinfo{author}{\bibfnamefont{B.}~\bibnamefont{Fr\"ohlich}},
  \bibinfo{author}{\bibfnamefont{M.}~\bibnamefont{Fattori}},
  \bibinfo{author}{\bibfnamefont{J.}~\bibnamefont{Metz}},
  \bibinfo{author}{\bibfnamefont{A.}~\bibnamefont{Griesmaier}},
  \bibinfo{author}{\bibfnamefont{S.}~\bibnamefont{Giovanazzi}},
  \bibnamefont{and} \bibinfo{author}{\bibfnamefont{T.}~\bibnamefont{Pfau}},
  \bibinfo{journal}{Nature} \textbf{\bibinfo{volume}{448}},
  \bibinfo{pages}{672} (\bibinfo{year}{2007}).

\bibitem[{\citenamefont{Lahaye et~al.}(2009)\citenamefont{Lahaye, Menotti,
  Santos, Lewenstein, and Pfau}}]{Lahaye09}
\bibinfo{author}{\bibfnamefont{T.}~\bibnamefont{Lahaye}},
  \bibinfo{author}{\bibfnamefont{C.}~\bibnamefont{Menotti}},
  \bibinfo{author}{\bibfnamefont{L.}~\bibnamefont{Santos}},
  \bibinfo{author}{\bibfnamefont{M.}~\bibnamefont{Lewenstein}},
  \bibnamefont{and} \bibinfo{author}{\bibfnamefont{T.}~\bibnamefont{Pfau}},
  \bibinfo{journal}{Rep. Prog. Phys.} \textbf{\bibinfo{volume}{72}},
  \bibinfo{pages}{126401} (\bibinfo{year}{2009}).

\bibitem[{\citenamefont{Lu et~al.}(2011)\citenamefont{Lu, Burdick, Youn, and
  Lev}}]{Lu11}
\bibinfo{author}{\bibfnamefont{M.}~\bibnamefont{Lu}},
  \bibinfo{author}{\bibfnamefont{N.~Q.} \bibnamefont{Burdick}},
  \bibinfo{author}{\bibfnamefont{S.~H.} \bibnamefont{Youn}}, \bibnamefont{and}
  \bibinfo{author}{\bibfnamefont{B.~L.} \bibnamefont{Lev}},
  \bibinfo{journal}{Phys. Rev. Lett.} \textbf{\bibinfo{volume}{107}},
  \bibinfo{pages}{190401} (\bibinfo{year}{2011}).

\bibitem[{\citenamefont{Lu et~al.}(2012)\citenamefont{Lu, Burdick, and
  Lev}}]{Lu12}
\bibinfo{author}{\bibfnamefont{M.}~\bibnamefont{Lu}},
  \bibinfo{author}{\bibfnamefont{N.~Q.} \bibnamefont{Burdick}},
  \bibnamefont{and} \bibinfo{author}{\bibfnamefont{B.~L.} \bibnamefont{Lev}},
  \bibinfo{journal}{Phys. Rev. Lett.} \textbf{\bibinfo{volume}{108}},
  \bibinfo{pages}{215301} (\bibinfo{year}{2012}).

\bibitem[{\citenamefont{Aikawa et~al.}(2012)\citenamefont{Aikawa, Frisch, Mark,
  Baier, Rietzler, Grimm, and Ferlaino}}]{Aikawa12}
\bibinfo{author}{\bibfnamefont{K.}~\bibnamefont{Aikawa}},
  \bibinfo{author}{\bibfnamefont{A.}~\bibnamefont{Frisch}},
  \bibinfo{author}{\bibfnamefont{M.}~\bibnamefont{Mark}},
  \bibinfo{author}{\bibfnamefont{S.}~\bibnamefont{Baier}},
  \bibinfo{author}{\bibfnamefont{A.}~\bibnamefont{Rietzler}},
  \bibinfo{author}{\bibfnamefont{R.}~\bibnamefont{Grimm}}, \bibnamefont{and}
  \bibinfo{author}{\bibfnamefont{F.}~\bibnamefont{Ferlaino}},
  \bibinfo{journal}{Phys. Rev. Lett.} \textbf{\bibinfo{volume}{108}},
  \bibinfo{pages}{210401} (\bibinfo{year}{2012}).

\bibitem[{\citenamefont{Bismut et~al.}(2012)\citenamefont{Bismut,
  Laburthe-Tolra, Mar\'echal, Pedri, Gorceix, and Vernac}}]{Bismut12}
\bibinfo{author}{\bibfnamefont{G.}~\bibnamefont{Bismut}},
  \bibinfo{author}{\bibfnamefont{B.}~\bibnamefont{Laburthe-Tolra}},
  \bibinfo{author}{\bibfnamefont{E.}~\bibnamefont{Mar\'echal}},
  \bibinfo{author}{\bibfnamefont{P.}~\bibnamefont{Pedri}},
  \bibinfo{author}{\bibfnamefont{O.}~\bibnamefont{Gorceix}}, \bibnamefont{and}
  \bibinfo{author}{\bibfnamefont{L.}~\bibnamefont{Vernac}},
  \bibinfo{journal}{Phys. Rev. Lett.} \textbf{\bibinfo{volume}{109}},
  \bibinfo{pages}{155302} (\bibinfo{year}{2012}).

\bibitem[{\citenamefont{Baranov et~al.}(2012)\citenamefont{Baranov, Dalmonte,
  Pupillo, and Zoller}}]{Baranov12}
\bibinfo{author}{\bibfnamefont{M.~A.} \bibnamefont{Baranov}},
  \bibinfo{author}{\bibfnamefont{M.}~\bibnamefont{Dalmonte}},
  \bibinfo{author}{\bibfnamefont{G.}~\bibnamefont{Pupillo}}, \bibnamefont{and}
  \bibinfo{author}{\bibfnamefont{P.}~\bibnamefont{Zoller}},
  \bibinfo{journal}{Chemical Reviews} \textbf{\bibinfo{volume}{112}},
  \bibinfo{pages}{5012} (\bibinfo{year}{2012}).

\bibitem[{\citenamefont{Ni et~al.}(2008)\citenamefont{Ni, Ospelkaus, {de
  Miranda}, Pe'er, Neyenhuis, Zirbel, Kotochigova, Julienne, Jin, and
  Ye}}]{Ni08}
\bibinfo{author}{\bibfnamefont{K.-K.} \bibnamefont{Ni}},
  \bibinfo{author}{\bibfnamefont{S.}~\bibnamefont{Ospelkaus}},
  \bibinfo{author}{\bibfnamefont{M.~H.~G.} \bibnamefont{{de Miranda}}},
  \bibinfo{author}{\bibfnamefont{A.}~\bibnamefont{Pe'er}},
  \bibinfo{author}{\bibfnamefont{B.}~\bibnamefont{Neyenhuis}},
  \bibinfo{author}{\bibfnamefont{J.~J.} \bibnamefont{Zirbel}},
  \bibinfo{author}{\bibfnamefont{S.}~\bibnamefont{Kotochigova}},
  \bibinfo{author}{\bibfnamefont{P.~S.} \bibnamefont{Julienne}},
  \bibinfo{author}{\bibfnamefont{D.~S.} \bibnamefont{Jin}}, \bibnamefont{and}
  \bibinfo{author}{\bibfnamefont{J.}~\bibnamefont{Ye}},
  \bibinfo{journal}{Science} \textbf{\bibinfo{volume}{322}},
  \bibinfo{pages}{231} (\bibinfo{year}{2008}).

\bibitem[{\citenamefont{Daiglmayr et~al.}(2008)\citenamefont{Daiglmayr,
  Grochola, Repp, M\"ortlbauer, Gl\"uck, Lange, Dulieu, Wester, and
  Weidem\"uller}}]{Deiglmayr08}
\bibinfo{author}{\bibfnamefont{J.}~\bibnamefont{Daiglmayr}},
  \bibinfo{author}{\bibfnamefont{A.}~\bibnamefont{Grochola}},
  \bibinfo{author}{\bibfnamefont{M.}~\bibnamefont{Repp}},
  \bibinfo{author}{\bibfnamefont{K.}~\bibnamefont{M\"ortlbauer}},
  \bibinfo{author}{\bibfnamefont{C.}~\bibnamefont{Gl\"uck}},
  \bibinfo{author}{\bibfnamefont{J.}~\bibnamefont{Lange}},
  \bibinfo{author}{\bibfnamefont{O.}~\bibnamefont{Dulieu}},
  \bibinfo{author}{\bibfnamefont{R.}~\bibnamefont{Wester}}, \bibnamefont{and}
  \bibinfo{author}{\bibfnamefont{M.}~\bibnamefont{Weidem\"uller}},
  \bibinfo{journal}{Phys. Rev. Lett.} \textbf{\bibinfo{volume}{101}},
  \bibinfo{pages}{133004} (\bibinfo{year}{2008}).

\bibitem[{\citenamefont{Aikawa et~al.}(2009)\citenamefont{Aikawa, Akamatsu,
  Kobayashi, Ueda, Kishimoto, and Inouye}}]{Aikawa09}
\bibinfo{author}{\bibfnamefont{K.}~\bibnamefont{Aikawa}},
  \bibinfo{author}{\bibfnamefont{D.}~\bibnamefont{Akamatsu}},
  \bibinfo{author}{\bibfnamefont{J.}~\bibnamefont{Kobayashi}},
  \bibinfo{author}{\bibfnamefont{M.}~\bibnamefont{Ueda}},
  \bibinfo{author}{\bibfnamefont{T.}~\bibnamefont{Kishimoto}},
  \bibnamefont{and} \bibinfo{author}{\bibfnamefont{S.}~\bibnamefont{Inouye}},
  \bibinfo{journal}{New J. Phys.} \textbf{\bibinfo{volume}{11}},
  \bibinfo{pages}{055035} (\bibinfo{year}{2009}).

\bibitem[{\citenamefont{Carr et~al.}(2009)\citenamefont{Carr, Demille, Krems,
  and Ye}}]{Carr09}
\bibinfo{author}{\bibfnamefont{L.~D.} \bibnamefont{Carr}},
  \bibinfo{author}{\bibfnamefont{D.}~\bibnamefont{Demille}},
  \bibinfo{author}{\bibfnamefont{R.}~\bibnamefont{Krems}}, \bibnamefont{and}
  \bibinfo{author}{\bibfnamefont{J.}~\bibnamefont{Ye}}, \bibinfo{journal}{New
  J. Phys.} \textbf{\bibinfo{volume}{11}}, \bibinfo{pages}{055049}
  (\bibinfo{year}{2009}).

\bibitem[{\citenamefont{Takekoshi et~al.}(2014)\citenamefont{Takekoshi,
  Reichs\"ollner, Schindewolf, Hutson, Sueur, Dulieu, Ferlaino, Grimm, and
  N\"agerl}}]{Takekoshi14}
\bibinfo{author}{\bibfnamefont{T.}~\bibnamefont{Takekoshi}},
  \bibinfo{author}{\bibfnamefont{L.}~\bibnamefont{Reichs\"ollner}},
  \bibinfo{author}{\bibfnamefont{A.}~\bibnamefont{Schindewolf}},
  \bibinfo{author}{\bibfnamefont{J.~M.} \bibnamefont{Hutson}},
  \bibinfo{author}{\bibfnamefont{C.~R.~L.} \bibnamefont{Sueur}},
  \bibinfo{author}{\bibfnamefont{O.}~\bibnamefont{Dulieu}},
  \bibinfo{author}{\bibfnamefont{F.}~\bibnamefont{Ferlaino}},
  \bibinfo{author}{\bibfnamefont{R.}~\bibnamefont{Grimm}}, \bibnamefont{and}
  \bibinfo{author}{\bibfnamefont{H.-C.} \bibnamefont{N\"agerl}},
  \bibinfo{journal}{Phys. Rev. Lett.} \textbf{\bibinfo{volume}{113}},
  \bibinfo{pages}{205301} (\bibinfo{year}{2014}).

\bibitem[{\citenamefont{G\'oral et~al.}(2002)\citenamefont{G\'oral, Santos, and
  Lewenstein}}]{Goral02}
\bibinfo{author}{\bibfnamefont{K.}~\bibnamefont{G\'oral}},
  \bibinfo{author}{\bibfnamefont{L.}~\bibnamefont{Santos}}, \bibnamefont{and}
  \bibinfo{author}{\bibfnamefont{M.}~\bibnamefont{Lewenstein}},
  \bibinfo{journal}{Phys. Rev. Lett.} \textbf{\bibinfo{volume}{88}},
  \bibinfo{pages}{170406} (\bibinfo{year}{2002}).

\bibitem[{\citenamefont{Kovrizhin et~al.}(2005)\citenamefont{Kovrizhin, Pai,
  and Sinha}}]{Kovrizhin05}
\bibinfo{author}{\bibfnamefont{D.~L.} \bibnamefont{Kovrizhin}},
  \bibinfo{author}{\bibfnamefont{G.~V.} \bibnamefont{Pai}}, \bibnamefont{and}
  \bibinfo{author}{\bibfnamefont{S.}~\bibnamefont{Sinha}},
  \bibinfo{journal}{Europhys. Lett.} \textbf{\bibinfo{volume}{72}},
  \bibinfo{pages}{162} (\bibinfo{year}{2005}).

\bibitem[{\citenamefont{Scarola and {Das Sarma}}(2005)}]{Scarola05}
\bibinfo{author}{\bibfnamefont{V.~W.} \bibnamefont{Scarola}} \bibnamefont{and}
  \bibinfo{author}{\bibfnamefont{S.}~\bibnamefont{{Das Sarma}}},
  \bibinfo{journal}{Phys. Rev. Lett.} \textbf{\bibinfo{volume}{95}},
  \bibinfo{pages}{033003} (\bibinfo{year}{2005}).

\bibitem[{\citenamefont{Yi et~al.}(2007)\citenamefont{Yi, Li, and Sun}}]{Yi07}
\bibinfo{author}{\bibfnamefont{S.}~\bibnamefont{Yi}},
  \bibinfo{author}{\bibfnamefont{T.}~\bibnamefont{Li}}, \bibnamefont{and}
  \bibinfo{author}{\bibfnamefont{C.~P.} \bibnamefont{Sun}},
  \bibinfo{journal}{Phys. Rev. Lett.} \textbf{\bibinfo{volume}{98}},
  \bibinfo{pages}{260405} (\bibinfo{year}{2007}).

\bibitem[{\citenamefont{Menotti et~al.}(2007)\citenamefont{Menotti, Trefzger,
  and Lewenstein}}]{Menotti07}
\bibinfo{author}{\bibfnamefont{C.}~\bibnamefont{Menotti}},
  \bibinfo{author}{\bibfnamefont{C.}~\bibnamefont{Trefzger}}, \bibnamefont{and}
  \bibinfo{author}{\bibfnamefont{M.}~\bibnamefont{Lewenstein}},
  \bibinfo{journal}{Phys. Rev. Lett.} \textbf{\bibinfo{volume}{98}},
  \bibinfo{pages}{235301} (\bibinfo{year}{2007}).

\bibitem[{\citenamefont{Danshita and {S\'a de Melo}}(2009)}]{Danshita09}
\bibinfo{author}{\bibfnamefont{I.}~\bibnamefont{Danshita}} \bibnamefont{and}
  \bibinfo{author}{\bibfnamefont{C.~A.~R.} \bibnamefont{{S\'a de Melo}}},
  \bibinfo{journal}{Phys. Rev. Lett.} \textbf{\bibinfo{volume}{103}},
  \bibinfo{pages}{225301} (\bibinfo{year}{2009}).

\bibitem[{\citenamefont{Burnell et~al.}(2009)\citenamefont{Burnell, Parish,
  Cooper, and Sondhi}}]{Burnell09}
\bibinfo{author}{\bibfnamefont{F.~J.} \bibnamefont{Burnell}},
  \bibinfo{author}{\bibfnamefont{M.~M.} \bibnamefont{Parish}},
  \bibinfo{author}{\bibfnamefont{N.~R.} \bibnamefont{Cooper}},
  \bibnamefont{and} \bibinfo{author}{\bibfnamefont{S.~L.}
  \bibnamefont{Sondhi}}, \bibinfo{journal}{Phys. Rev. B}
  \textbf{\bibinfo{volume}{80}}, \bibinfo{pages}{174519}
  (\bibinfo{year}{2009}).

\bibitem[{\citenamefont{Cinti et~al.}(2010)\citenamefont{Cinti, Jain,
  Bononsegni, Micheli, Zoller, and Pupillo}}]{Cinto10}
\bibinfo{author}{\bibfnamefont{F.}~\bibnamefont{Cinti}},
  \bibinfo{author}{\bibfnamefont{P.}~\bibnamefont{Jain}},
  \bibinfo{author}{\bibfnamefont{M.}~\bibnamefont{Bononsegni}},
  \bibinfo{author}{\bibfnamefont{A.}~\bibnamefont{Micheli}},
  \bibinfo{author}{\bibfnamefont{P.}~\bibnamefont{Zoller}}, \bibnamefont{and}
  \bibinfo{author}{\bibfnamefont{G.}~\bibnamefont{Pupillo}},
  \bibinfo{journal}{Phys. Rev. Lett.} \textbf{\bibinfo{volume}{105}},
  \bibinfo{pages}{135301} (\bibinfo{year}{2010}).

\bibitem[{\citenamefont{Trefzger et~al.}(2010)\citenamefont{Trefzger, Alloing,
  Menotti, Dubin, and Lewenstein}}]{Trefzger10}
\bibinfo{author}{\bibfnamefont{C.}~\bibnamefont{Trefzger}},
  \bibinfo{author}{\bibfnamefont{M.}~\bibnamefont{Alloing}},
  \bibinfo{author}{\bibfnamefont{C.}~\bibnamefont{Menotti}},
  \bibinfo{author}{\bibfnamefont{F.}~\bibnamefont{Dubin}}, \bibnamefont{and}
  \bibinfo{author}{\bibfnamefont{M.}~\bibnamefont{Lewenstein}},
  \bibinfo{journal}{New J. Phys.} \textbf{\bibinfo{volume}{12}},
  \bibinfo{pages}{093008} (\bibinfo{year}{2010}).

\bibitem[{\citenamefont{Pollet et~al.}(2010)\citenamefont{Pollet, Picon,
  B\"uchler, and Troyer}}]{Pollet10}
\bibinfo{author}{\bibfnamefont{L.}~\bibnamefont{Pollet}},
  \bibinfo{author}{\bibfnamefont{J.~D.} \bibnamefont{Picon}},
  \bibinfo{author}{\bibfnamefont{H.~P.} \bibnamefont{B\"uchler}},
  \bibnamefont{and} \bibinfo{author}{\bibfnamefont{M.}~\bibnamefont{Troyer}},
  \bibinfo{journal}{Phys. Rev. Lett.} \textbf{\bibinfo{volume}{104}},
  \bibinfo{pages}{125302} (\bibinfo{year}{2010}).

\bibitem[{\citenamefont{Iskin}(2011)}]{Iskin11}
\bibinfo{author}{\bibfnamefont{M.}~\bibnamefont{Iskin}},
  \bibinfo{journal}{Phys. Rev. A} \textbf{\bibinfo{volume}{83}},
  \bibinfo{pages}{051606(R)} (\bibinfo{year}{2011}).

\bibitem[{\citenamefont{Trefzger et~al.}(2011)\citenamefont{Trefzger, Menotti,
  Capogrosso-Sansone, and Lewenstein}}]{Trefzger11}
\bibinfo{author}{\bibfnamefont{C.}~\bibnamefont{Trefzger}},
  \bibinfo{author}{\bibfnamefont{C.}~\bibnamefont{Menotti}},
  \bibinfo{author}{\bibfnamefont{B.}~\bibnamefont{Capogrosso-Sansone}},
  \bibnamefont{and}
  \bibinfo{author}{\bibfnamefont{M.}~\bibnamefont{Lewenstein}},
  \bibinfo{journal}{J. Phys. B} \textbf{\bibinfo{volume}{44}},
  \bibinfo{pages}{193001} (\bibinfo{year}{2011}).

\bibitem[{\citenamefont{Fellows and Carr}(2011)}]{Fellows11}
\bibinfo{author}{\bibfnamefont{J.~M.} \bibnamefont{Fellows}} \bibnamefont{and}
  \bibinfo{author}{\bibfnamefont{S.~T.} \bibnamefont{Carr}},
  \bibinfo{journal}{Phys. Rev. A} \textbf{\bibinfo{volume}{84}},
  \bibinfo{pages}{051602(R)} (\bibinfo{year}{2011}).

\bibitem[{\citenamefont{Ohgoe et~al.}(2012)\citenamefont{Ohgoe, Suzuki, and
  Kawashima}}]{Ohgoe12}
\bibinfo{author}{\bibfnamefont{T.}~\bibnamefont{Ohgoe}},
  \bibinfo{author}{\bibfnamefont{T.}~\bibnamefont{Suzuki}}, \bibnamefont{and}
  \bibinfo{author}{\bibfnamefont{N.}~\bibnamefont{Kawashima}},
  \bibinfo{journal}{Phys. Rev. B} \textbf{\bibinfo{volume}{86}},
  \bibinfo{pages}{054520} (\bibinfo{year}{2012}).

\bibitem[{\citenamefont{Yamamoto et~al.}(2012)\citenamefont{Yamamoto, Danshita,
  and {S\'a de Melo}}}]{Yamamoto12}
\bibinfo{author}{\bibfnamefont{D.}~\bibnamefont{Yamamoto}},
  \bibinfo{author}{\bibfnamefont{I.}~\bibnamefont{Danshita}}, \bibnamefont{and}
  \bibinfo{author}{\bibfnamefont{C.~A.~R.} \bibnamefont{{S\'a de Melo}}},
  \bibinfo{journal}{Phys. Rev. A} \textbf{\bibinfo{volume}{85}},
  \bibinfo{pages}{021601(R)} (\bibinfo{year}{2012}).

\bibitem[{\citenamefont{Lu et~al.}(2015)\citenamefont{Lu, Li, Petrov, and
  Shlyapnikov}}]{Lu15}
\bibinfo{author}{\bibfnamefont{Z.-K.} \bibnamefont{Lu}},
  \bibinfo{author}{\bibfnamefont{Y.}~\bibnamefont{Li}},
  \bibinfo{author}{\bibfnamefont{D.~S.} \bibnamefont{Petrov}},
  \bibnamefont{and} \bibinfo{author}{\bibfnamefont{G.~V.}
  \bibnamefont{Shlyapnikov}}, \bibinfo{journal}{Phys. Rev. Lett.}
  \textbf{\bibinfo{volume}{115}}, \bibinfo{pages}{075303}
  (\bibinfo{year}{2015}).

\bibitem[{\citenamefont{Chester}(1970)}]{Chester70}
\bibinfo{author}{\bibfnamefont{G.~V.} \bibnamefont{Chester}},
  \bibinfo{journal}{Phys. Rev. A} \textbf{\bibinfo{volume}{2}},
  \bibinfo{pages}{256} (\bibinfo{year}{1970}).

\bibitem[{\citenamefont{Leggett}(1970)}]{Leggett70}
\bibinfo{author}{\bibfnamefont{A.~J.} \bibnamefont{Leggett}},
  \bibinfo{journal}{Phys. Rev. Lett.} \textbf{\bibinfo{volume}{25}},
  \bibinfo{pages}{1543} (\bibinfo{year}{1970}).

\bibitem[{\citenamefont{Kim and Chan}(2004)}]{Kim04}
\bibinfo{author}{\bibfnamefont{E.}~\bibnamefont{Kim}} \bibnamefont{and}
  \bibinfo{author}{\bibfnamefont{M.~H.~W.} \bibnamefont{Chan}},
  \bibinfo{journal}{Nature} \textbf{\bibinfo{volume}{427}},
  \bibinfo{pages}{225} (\bibinfo{year}{2004}).

\bibitem[{\citenamefont{Kim and Chan}(2006)}]{Kim06}
\bibinfo{author}{\bibfnamefont{E.}~\bibnamefont{Kim}} \bibnamefont{and}
  \bibinfo{author}{\bibfnamefont{M.~H.~W.} \bibnamefont{Chan}},
  \bibinfo{journal}{Phys. Rev. Lett.} \textbf{\bibinfo{volume}{97}},
  \bibinfo{pages}{115302} (\bibinfo{year}{2006}).

\bibitem[{\citenamefont{Todoshchenko et~al.}(2007)\citenamefont{Todoshchenko,
  Alles, Junes, Parshin, and Tsepelin}}]{Todoshchenko07}
\bibinfo{author}{\bibfnamefont{I.~A.} \bibnamefont{Todoshchenko}},
  \bibinfo{author}{\bibfnamefont{H.}~\bibnamefont{Alles}},
  \bibinfo{author}{\bibfnamefont{H.~J.} \bibnamefont{Junes}},
  \bibinfo{author}{\bibfnamefont{A.~Y.} \bibnamefont{Parshin}},
  \bibnamefont{and} \bibinfo{author}{\bibfnamefont{V.}~\bibnamefont{Tsepelin}},
  \bibinfo{journal}{JETP Lett.} \textbf{\bibinfo{volume}{85}},
  \bibinfo{pages}{454} (\bibinfo{year}{2007}).

\bibitem[{\citenamefont{Choi et~al.}(2010)\citenamefont{Choi, Takahashi, Kono,
  and Kim}}]{Choi10}
\bibinfo{author}{\bibfnamefont{H.}~\bibnamefont{Choi}},
  \bibinfo{author}{\bibfnamefont{D.}~\bibnamefont{Takahashi}},
  \bibinfo{author}{\bibfnamefont{K.}~\bibnamefont{Kono}}, \bibnamefont{and}
  \bibinfo{author}{\bibfnamefont{E.}~\bibnamefont{Kim}},
  \bibinfo{journal}{Science} \textbf{\bibinfo{volume}{330}},
  \bibinfo{pages}{1512} (\bibinfo{year}{2010}).

\bibitem[{\citenamefont{Kim and Chan}(2012)}]{Kim12}
\bibinfo{author}{\bibfnamefont{D.~Y.} \bibnamefont{Kim}} \bibnamefont{and}
  \bibinfo{author}{\bibfnamefont{M.~H.~W.} \bibnamefont{Chan}},
  \bibinfo{journal}{Phys. Rev. Lett.} \textbf{\bibinfo{volume}{109}},
  \bibinfo{pages}{155301} (\bibinfo{year}{2012}).

\bibitem[{\citenamefont{van Otterlo and Wagenblast}(1994)}]{Otterlo94}
\bibinfo{author}{\bibfnamefont{A.}~\bibnamefont{van Otterlo}} \bibnamefont{and}
  \bibinfo{author}{\bibfnamefont{K.-H.} \bibnamefont{Wagenblast}},
  \bibinfo{journal}{Phys. Rev. Lett.} \textbf{\bibinfo{volume}{72}},
  \bibinfo{pages}{3598} (\bibinfo{year}{1994}).

\bibitem[{\citenamefont{van Otterlo et~al.}(1995)\citenamefont{van Otterlo,
  Wagenblast, Baltin, Bruder, Fazio, and Sch\"on}}]{Otterlo95}
\bibinfo{author}{\bibfnamefont{A.}~\bibnamefont{van Otterlo}},
  \bibinfo{author}{\bibfnamefont{K.-H.} \bibnamefont{Wagenblast}},
  \bibinfo{author}{\bibfnamefont{R.}~\bibnamefont{Baltin}},
  \bibinfo{author}{\bibfnamefont{C.}~\bibnamefont{Bruder}},
  \bibinfo{author}{\bibfnamefont{R.}~\bibnamefont{Fazio}}, \bibnamefont{and}
  \bibinfo{author}{\bibfnamefont{G.}~\bibnamefont{Sch\"on}},
  \bibinfo{journal}{Phys. Rev. B} \textbf{\bibinfo{volume}{52}},
  \bibinfo{pages}{16176} (\bibinfo{year}{1995}).

\bibitem[{\citenamefont{Batrouni et~al.}(1995)\citenamefont{Batrouni,
  Scalettar, Zimanyi, and Kampf}}]{Batrouni95}
\bibinfo{author}{\bibfnamefont{G.~G.} \bibnamefont{Batrouni}},
  \bibinfo{author}{\bibfnamefont{R.~T.} \bibnamefont{Scalettar}},
  \bibinfo{author}{\bibfnamefont{G.~T.} \bibnamefont{Zimanyi}},
  \bibnamefont{and} \bibinfo{author}{\bibfnamefont{A.~P.} \bibnamefont{Kampf}},
  \bibinfo{journal}{Phys. Rev. Lett.} \textbf{\bibinfo{volume}{74}},
  \bibinfo{pages}{2527} (\bibinfo{year}{1995}).

\bibitem[{\citenamefont{Baumann et~al.}(2010)\citenamefont{Baumann, Guerlin,
  Brennecke, and Esslinger}}]{Baumann10}
\bibinfo{author}{\bibfnamefont{K.}~\bibnamefont{Baumann}},
  \bibinfo{author}{\bibfnamefont{C.}~\bibnamefont{Guerlin}},
  \bibinfo{author}{\bibfnamefont{F.}~\bibnamefont{Brennecke}},
  \bibnamefont{and}
  \bibinfo{author}{\bibfnamefont{T.}~\bibnamefont{Esslinger}},
  \bibinfo{journal}{Nature} \textbf{\bibinfo{volume}{464}},
  \bibinfo{pages}{1301} (\bibinfo{year}{2010}).

\bibitem[{\citenamefont{B\"uhler and B\"uchler}(2011)}]{Buhler11}
\bibinfo{author}{\bibfnamefont{A.}~\bibnamefont{B\"uhler}} \bibnamefont{and}
  \bibinfo{author}{\bibfnamefont{H.~P.} \bibnamefont{B\"uchler}},
  \bibinfo{journal}{Phys. Rev. A} \textbf{\bibinfo{volume}{84}},
  \bibinfo{pages}{023607} (\bibinfo{year}{2011}).

\bibitem[{\citenamefont{Wilson et~al.}(2012)\citenamefont{Wilson, Ticknor,
  Bohn, and Timmermans}}]{Wilson12}
\bibinfo{author}{\bibfnamefont{R.~M.} \bibnamefont{Wilson}},
  \bibinfo{author}{\bibfnamefont{C.}~\bibnamefont{Ticknor}},
  \bibinfo{author}{\bibfnamefont{J.~L.} \bibnamefont{Bohn}}, \bibnamefont{and}
  \bibinfo{author}{\bibfnamefont{E.}~\bibnamefont{Timmermans}},
  \bibinfo{journal}{Phys. Rev. A} \textbf{\bibinfo{volume}{86}},
  \bibinfo{pages}{033606} (\bibinfo{year}{2012}).

\bibitem[{\citenamefont{Shirley et~al.}(2014)\citenamefont{Shirley, Anderson,
  Clark, and Wilson}}]{Shirley14}
\bibinfo{author}{\bibfnamefont{W.~E.} \bibnamefont{Shirley}},
  \bibinfo{author}{\bibfnamefont{B.~M.} \bibnamefont{Anderson}},
  \bibinfo{author}{\bibfnamefont{C.~W.} \bibnamefont{Clark}}, \bibnamefont{and}
  \bibinfo{author}{\bibfnamefont{R.~M.} \bibnamefont{Wilson}},
  \bibinfo{journal}{Phys. Rev. Lett.} \textbf{\bibinfo{volume}{113}},
  \bibinfo{pages}{165301} (\bibinfo{year}{2014}).

\bibitem[{\citenamefont{Rokhsar and Kotliar}(1991)}]{Rokhsar91}
\bibinfo{author}{\bibfnamefont{D.~S.} \bibnamefont{Rokhsar}} \bibnamefont{and}
  \bibinfo{author}{\bibfnamefont{B.~G.} \bibnamefont{Kotliar}},
  \bibinfo{journal}{Phys. Rev. B} \textbf{\bibinfo{volume}{44}},
  \bibinfo{pages}{10328} (\bibinfo{year}{1991}).

\bibitem[{\citenamefont{Baier et~al.}(2015)\citenamefont{Baier, Mark, Petter,
  Aikawa, Chomaz, Cai, Baranov, Zoller, and Ferlaino}}]{BaierarXiv15}
\bibinfo{author}{\bibfnamefont{S.}~\bibnamefont{Baier}},
  \bibinfo{author}{\bibfnamefont{M.~J.} \bibnamefont{Mark}},
  \bibinfo{author}{\bibfnamefont{D.}~\bibnamefont{Petter}},
  \bibinfo{author}{\bibfnamefont{K.}~\bibnamefont{Aikawa}},
  \bibinfo{author}{\bibfnamefont{L.}~\bibnamefont{Chomaz}},
  \bibinfo{author}{\bibfnamefont{Z.}~\bibnamefont{Cai}},
  \bibinfo{author}{\bibfnamefont{M.}~\bibnamefont{Baranov}},
  \bibinfo{author}{\bibfnamefont{P.}~\bibnamefont{Zoller}}, \bibnamefont{and}
  \bibinfo{author}{\bibfnamefont{F.}~\bibnamefont{Ferlaino}}
  (\bibinfo{year}{2015}), \eprint{arxiv:1507.03500}.

\bibitem[{\citenamefont{Ospelkaus et~al.}(2010)\citenamefont{Ospelkaus, Ni,
  Wang, {de Miranda}, Neyenhuis, Qu\'em\'ener, Julienne, Bohn, Jin, and
  Ye}}]{Ospelkaus10}
\bibinfo{author}{\bibfnamefont{S.}~\bibnamefont{Ospelkaus}},
  \bibinfo{author}{\bibfnamefont{K.-K.} \bibnamefont{Ni}},
  \bibinfo{author}{\bibfnamefont{D.}~\bibnamefont{Wang}},
  \bibinfo{author}{\bibfnamefont{M.~H.~G.} \bibnamefont{{de Miranda}}},
  \bibinfo{author}{\bibfnamefont{B.}~\bibnamefont{Neyenhuis}},
  \bibinfo{author}{\bibfnamefont{G.}~\bibnamefont{Qu\'em\'ener}},
  \bibinfo{author}{\bibfnamefont{P.~S.} \bibnamefont{Julienne}},
  \bibinfo{author}{\bibfnamefont{J.~L.} \bibnamefont{Bohn}},
  \bibinfo{author}{\bibfnamefont{D.~S.} \bibnamefont{Jin}}, \bibnamefont{and}
  \bibinfo{author}{\bibfnamefont{J.}~\bibnamefont{Ye}},
  \bibinfo{journal}{Science} \textbf{\bibinfo{volume}{327}},
  \bibinfo{pages}{853} (\bibinfo{year}{2010}).

\bibitem[{\citenamefont{Ni et~al.}(2010)\citenamefont{Ni, Ospelkaus, Wang,
  Qu\'em\'ener, Neyenhuis, {de Miranda}, Bohn, Ye, and Jin}}]{Ni10}
\bibinfo{author}{\bibfnamefont{K.-K.} \bibnamefont{Ni}},
  \bibinfo{author}{\bibfnamefont{S.}~\bibnamefont{Ospelkaus}},
  \bibinfo{author}{\bibfnamefont{D.}~\bibnamefont{Wang}},
  \bibinfo{author}{\bibfnamefont{G.}~\bibnamefont{Qu\'em\'ener}},
  \bibinfo{author}{\bibfnamefont{B.}~\bibnamefont{Neyenhuis}},
  \bibinfo{author}{\bibfnamefont{M.~H.~G.} \bibnamefont{{de Miranda}}},
  \bibinfo{author}{\bibfnamefont{J.~L.} \bibnamefont{Bohn}},
  \bibinfo{author}{\bibfnamefont{J.}~\bibnamefont{Ye}}, \bibnamefont{and}
  \bibinfo{author}{\bibfnamefont{D.~S.} \bibnamefont{Jin}},
  \bibinfo{journal}{Nature} \textbf{\bibinfo{volume}{464}},
  \bibinfo{pages}{1324} (\bibinfo{year}{2010}).

\bibitem[{\citenamefont{Mayle et~al.}(2013)\citenamefont{Mayle, Qu\'em\'ener,
  Ruzic, and Bohn}}]{Mayle13}
\bibinfo{author}{\bibfnamefont{M.}~\bibnamefont{Mayle}},
  \bibinfo{author}{\bibfnamefont{G.}~\bibnamefont{Qu\'em\'ener}},
  \bibinfo{author}{\bibfnamefont{B.~P.} \bibnamefont{Ruzic}}, \bibnamefont{and}
  \bibinfo{author}{\bibfnamefont{J.~L.} \bibnamefont{Bohn}},
  \bibinfo{journal}{Phys. Rev. A} \textbf{\bibinfo{volume}{87}},
  \bibinfo{pages}{012709} (\bibinfo{year}{2013}).

\bibitem[{\citenamefont{Kuklov et~al.}(2004)\citenamefont{Kuklov, Prokof'ev,
  and Svistunov}}]{Kuklov04}
\bibinfo{author}{\bibfnamefont{A.}~\bibnamefont{Kuklov}},
  \bibinfo{author}{\bibfnamefont{N.}~\bibnamefont{Prokof'ev}},
  \bibnamefont{and}
  \bibinfo{author}{\bibfnamefont{B.}~\bibnamefont{Svistunov}},
  \bibinfo{journal}{Phys. Rev. Lett.} \textbf{\bibinfo{volume}{92}},
  \bibinfo{pages}{050402} (\bibinfo{year}{2004}).

\bibitem[{\citenamefont{Isacsson et~al.}(2005)\citenamefont{Isacsson, Cha,
  Sengupta, and Girvin}}]{Isacsson05}
\bibinfo{author}{\bibfnamefont{A.}~\bibnamefont{Isacsson}},
  \bibinfo{author}{\bibfnamefont{M.-C.} \bibnamefont{Cha}},
  \bibinfo{author}{\bibfnamefont{K.}~\bibnamefont{Sengupta}}, \bibnamefont{and}
  \bibinfo{author}{\bibfnamefont{S.~M.} \bibnamefont{Girvin}},
  \bibinfo{journal}{Phys. Rev. B} \textbf{\bibinfo{volume}{72}},
  \bibinfo{pages}{184507} (\bibinfo{year}{2005}).

\bibitem[{\citenamefont{Yamamoto et~al.}(2013)\citenamefont{Yamamoto, Ozaki,
  {S\'a de Melo}, and Danshita}}]{Yamamoto13}
\bibinfo{author}{\bibfnamefont{D.}~\bibnamefont{Yamamoto}},
  \bibinfo{author}{\bibfnamefont{T.}~\bibnamefont{Ozaki}},
  \bibinfo{author}{\bibfnamefont{C.~A.~R.} \bibnamefont{{S\'a de Melo}}},
  \bibnamefont{and} \bibinfo{author}{\bibfnamefont{I.}~\bibnamefont{Danshita}},
  \bibinfo{journal}{Phys. Rev. A} \textbf{\bibinfo{volume}{88}},
  \bibinfo{pages}{033624} (\bibinfo{year}{2013}).

\bibitem[{\citenamefont{Stenger et~al.}(1999)\citenamefont{Stenger, Inouye,
  Chikkatur, Stamper-Kurn, Prichard, and Ketterle}}]{Stenger99}
\bibinfo{author}{\bibfnamefont{J.}~\bibnamefont{Stenger}},
  \bibinfo{author}{\bibfnamefont{S.}~\bibnamefont{Inouye}},
  \bibinfo{author}{\bibfnamefont{A.~P.} \bibnamefont{Chikkatur}},
  \bibinfo{author}{\bibfnamefont{D.~M.} \bibnamefont{Stamper-Kurn}},
  \bibinfo{author}{\bibfnamefont{D.~E.} \bibnamefont{Prichard}},
  \bibnamefont{and} \bibinfo{author}{\bibfnamefont{W.}~\bibnamefont{Ketterle}},
  \bibinfo{journal}{Phys. Rev. Lett.} \textbf{\bibinfo{volume}{82}},
  \bibinfo{pages}{4569} (\bibinfo{year}{1999}).

\bibitem[{\citenamefont{Hart et~al.}(2015)\citenamefont{Hart, Duarte, Yang,
  Liu, Paiva, Khatami, Scalettar, Trivedi, Huse, and Hulet}}]{Hart15}
\bibinfo{author}{\bibfnamefont{R.~A.} \bibnamefont{Hart}},
  \bibinfo{author}{\bibfnamefont{P.~M.} \bibnamefont{Duarte}},
  \bibinfo{author}{\bibfnamefont{T.-L.} \bibnamefont{Yang}},
  \bibinfo{author}{\bibfnamefont{X.}~\bibnamefont{Liu}},
  \bibinfo{author}{\bibfnamefont{T.}~\bibnamefont{Paiva}},
  \bibinfo{author}{\bibfnamefont{E.}~\bibnamefont{Khatami}},
  \bibinfo{author}{\bibfnamefont{R.~T.} \bibnamefont{Scalettar}},
  \bibinfo{author}{\bibfnamefont{N.}~\bibnamefont{Trivedi}},
  \bibinfo{author}{\bibfnamefont{D.~A.} \bibnamefont{Huse}}, \bibnamefont{and}
  \bibinfo{author}{\bibfnamefont{R.~A.} \bibnamefont{Hulet}},
  \bibinfo{journal}{Nature} \textbf{\bibinfo{volume}{519}},
  \bibinfo{pages}{211} (\bibinfo{year}{2015}).

\bibitem[{\citenamefont{Capogrosso-Sansone
  et~al.}(2008)\citenamefont{Capogrosso-Sansone, S\"oyler, Prokof'ev, and
  Svistunov}}]{Capogrosso08}
\bibinfo{author}{\bibfnamefont{B.}~\bibnamefont{Capogrosso-Sansone}},
  \bibinfo{author}{\bibfnamefont{S.~G.} \bibnamefont{S\"oyler}},
  \bibinfo{author}{\bibfnamefont{N.}~\bibnamefont{Prokof'ev}},
  \bibnamefont{and}
  \bibinfo{author}{\bibfnamefont{B.}~\bibnamefont{Svistunov}},
  \bibinfo{journal}{Phys. Rev. A} \textbf{\bibinfo{volume}{77}},
  \bibinfo{pages}{015602} (\bibinfo{year}{2008}).

\bibitem[{\citenamefont{Capogrosso-Sansone
  et~al.}(2010)\citenamefont{Capogrosso-Sansone, Trefzger, Lewenstein, Zoller,
  and Pupillo}}]{Capogrosso10}
\bibinfo{author}{\bibfnamefont{B.}~\bibnamefont{Capogrosso-Sansone}},
  \bibinfo{author}{\bibfnamefont{C.}~\bibnamefont{Trefzger}},
  \bibinfo{author}{\bibfnamefont{M.}~\bibnamefont{Lewenstein}},
  \bibinfo{author}{\bibfnamefont{P.}~\bibnamefont{Zoller}}, \bibnamefont{and}
  \bibinfo{author}{\bibfnamefont{G.}~\bibnamefont{Pupillo}},
  \bibinfo{journal}{Phys. Rev. Lett.} \textbf{\bibinfo{volume}{104}},
  \bibinfo{pages}{125301} (\bibinfo{year}{2010}).

\bibitem[{\citenamefont{Kadau et~al.}(2015)\citenamefont{Kadau, Schmitt,
  Wenzel, Wink, Maier, Ferrier-Barbut, and Pfau}}]{Kadau15arXiv}
\bibinfo{author}{\bibfnamefont{H.}~\bibnamefont{Kadau}},
  \bibinfo{author}{\bibfnamefont{M.}~\bibnamefont{Schmitt}},
  \bibinfo{author}{\bibfnamefont{M.}~\bibnamefont{Wenzel}},
  \bibinfo{author}{\bibfnamefont{C.}~\bibnamefont{Wink}},
  \bibinfo{author}{\bibfnamefont{T.}~\bibnamefont{Maier}},
  \bibinfo{author}{\bibfnamefont{I.}~\bibnamefont{Ferrier-Barbut}},
  \bibnamefont{and} \bibinfo{author}{\bibfnamefont{T.}~\bibnamefont{Pfau}}
  (\bibinfo{year}{2015}), \eprint{arxiv:1508.05007}.

\end{thebibliography}
\end{document}